\documentclass{iopart}
\usepackage{amssymb,graphicx,enumerate,color}

\newcommand{\bignone}{}
\newcommand{\mathd}{\mathrm{d}}
\newcommand{\mathe}{\mathrm{e}}
\newcommand{\tmem}[1]{{\em #1\/}}
\newcommand{\tmtextit}[1]{{\itshape{#1}}}
\newcommand{\um}{-}
\newenvironment{enumeratealpha}{\begin{enumerate}[a{\textup{)}}] }{\end{enumerate}}
\newenvironment{enumeratenumeric}{\begin{enumerate}[1.] }{\end{enumerate}}

\begin{document}

\title{Universal features of cell polarization processes}
\author{A. Gamba$^{1,2,3}$,
I. Kolokolov$^4$,
V. Lebedev$^4$,
and G. Ortenzi$^1$
}
\address{$^1$ Politecnico di Torino and CNISM,
Corso Duca degli Abruzzi 24, Torino, Italy}
\address{$^2$ INFN, via Pietro Giuria 1, 10125 Torino, Italy}
\address{$^3$ Kavli Institute for Theoretical Physics, Santa Barbara,
CA 93106-4030, USA }
\address{$^4$ Landau Institute for Theoretical Physics, Kosygina 2, 119334
Moscow, Russia}
\ead{andrea.gamba@polito.it}

\begin{abstract}
  Cell polarization plays a central role in the development of complex
  organisms. It has been recently shown that cell polarization may follow from
  the proximity to a phase separation instability in a bistable network of
  chemical reactions. An example which has been thoroughly studied is the
  formation of signaling domains during eukaryotic chemotaxis. In this case,
  the process of domain growth may be described by the use of a constrained
  time-dependent Landau-Ginzburg equation, admitting scale-invariant solutions
  {\tmem{\`a la}} Lifshitz and Slyozov. The constraint results here from a
  mechanism of fast cycling of molecules between a cytosolic, inactive state
  and a membrane-bound, active state, which dynamically tunes the chemical
  potential for membrane binding to a value corresponding to the coexistence
  of different phases on the cell membrane. We provide here a universal
  description of this process both in the presence and absence of a gradient
  in the external activation field. Universal power laws are derived for the
  time needed for the cell to polarize in a chemotactic gradient, and for the
  value of the smallest detectable gradient. We also describe a concrete
  realization of our scheme based on the analysis of available biochemical and
  biophysical data.
\end{abstract}
 

\section{Introduction}

Biophysical processes of cell polarization have attracted large interest in
recent times. It has been observed that intriguing similarities exist in the
polarization of such diverse biological systems as cells of the immune system,
social amoebas, budding yeast, and amphibian eggs {\cite{WL03}}. This suggests
that cell polarization may be a highly universal phenomenon.

One of the best studied examples of the role of biochemical cell membrane
polarization in eukaryotic cells is chemotaxis. Chemotaxis is the ability of
cells to sense spatial gradients of attractant factors, governing the
development of all superior organisms. Eukaryotic cells are endowed with an
extremely sensible chemical compass allowing them to orient toward sources of
soluble chemical signals. This mechanism is the result of billion years of
evolution, and multicellular organisms would not exist without it. Slight
gradients in the external signals produced by the environment induce the
formation of oriented domains of signaling molecules on the cell membrane
surface. Afterwards, these signaling domains induce differentiated
polymerization of the cell cytoskeleton in their proximity, inducing the
formation of a growing head and a retracting tail, and eventually directed
motion towards the attractant source.

It has been suggested in the biological literature that domains of signaling
molecules are self-organized structures {\cite{PRG+04}}. In this paper we
confirm that this expectation may be substantiated by the use of statistical
mechanical methods, leading to the prediction that universal features typical
of coarsening processes in phase-ordering systems should be observable in
polarizing cells. We also describe here a concrete realization of our scheme
in the process of eukaryotic chemotaxis, based on the analysis of available
biochemical and biophysical data. Part of the results presented here have been
briefly reported in a previous letter {\cite{GKL+07}}.

\section{Cell polarity}

Stochastic reaction-diffusion systems are a natural paradigm for describing in
physical terms the biochemical processes taking place in the living cell,
since the cytosol and cell membrane are inherently diffusive
environments{\hspace{1pt}}{\footnote{For general facts regarding cell biology
we refer to Ref. {\cite{AJL+07}}.}}. Although active transport processes also
take place in the cell, they regard mainly vesicles, organelles and large
multiprotein complexes, while smaller cell constituents move diffusively.
Thermal agitation and the intrinsic stochasticity in the advancement of
chemical reactions provide natural sources of noise.

Most reactions in the cellular environment would be very slow if they were not
favored by the action of catalysts. Small numbers of enzymatic molecules
($10^3$--$10^5$ per cell) control the speed of chemical reactions involving
much larger numbers of substrate molecules ($10^5$--$10^6$ per cell.) Often,
the substrate concentration in its turn controls the catalyst activity, so
that the response of the system becomes nonlinear. Most biochemically relevant
reactions involve enzyme-substrate couples and are part of networks of
interconnected autocatalytic reactions.

Nonlinearities allow in principle the system to realize several stable
biochemical phases, characterized by different concentrations of chemical
factors {\cite{Kam07}}. Transitions between different phases in
reaction-diffusion systems have been observed in purely physical settings,
such as the adsorption and reaction of gases on catalytic surfaces
{\cite{SHV+01,WHS+05}}. Recently, it has been shown that a similar process of
nonequilibrium phase separation may be at the heart of directional sensing in
higher eukaryotes {\cite{GCT+05,GKL+07}}.

In eukaryotic directional sensing cells exposed to shallow gradients of
external attractant factors polarize accumulating the phospholipidic signaling
molecule {\tmem{phosphatidylinositol trisphosphate}} (PIP$3$) and the
PIP3-producing enzyme {\tmem{phosphatidylinositol 3-kinase}} (PI3K) on the
cell membrane side exposed to the highest attractant concentrations, while
{\tmem{phosphatidylinositol bisphosphate}} (PIP2) and the PIP2-producing
enzyme {\tmem{phosphatase and tensin homolog}} (PTEN) accumulate on the
complementary side{\hspace{1pt}} {\cite{PD99}} (see \ref{app:lattice}
for a more abstract description of the relative roles of these signaling
molecules.)

Accurate quantitative experiments {\cite{SNB+06,SMO06}} performed by exposing
{\tmem{Dictyostelium}} cells to controlled attractant gradients showed that
uniform concentrations of external attractant factor induce a predominant,
uniform concentration of PIP3 and PI3K on the cell membrane, and do not
immediately result in cell polarization and motion. However, slight gradients
in the distribution of the attractant factor induce the formation of two
complementary domains, one rich in PIP3 and PI3K, and one rich in PIP2 and
PTEN, in times of the order of a few minutes. This early breaking of the
spherical symmetry of the cell membrane induces cell polarization and motion
{\cite{PD99}}. Uniformly stimulated cells observed over longer timescales (of
the order of 1 hour) are seen to polarize stochastically and move in random
directions.

Numerical simulations of a stochastic reaction-diffusion model of the process
suggest that both the early, large amplification of slight attractant
gradients and the separate phenomenon of late, random polarization under
uniform stimulation are explained by the proximity of the system to a
spontaneous phase separation driven by non-linear autocatalytic interactions
{\cite{GCT+05}}. In this framework, cell polarization is the final result of a
nucleation process by which domains rich in PIP3 and PI3K are created in a sea
rich in PIP2 and PTEN, or vice versa, depending on initial conditions and
activation patterns. The polarization process is accomplished when pure PIP2
and PIP3 rich domains grow to sizes comparable to the size of the cell.
Gradient activation patterns strongly influence the kinetic of domain growth
and coalescence, taking advantage of the underlying phase-separation
instability. This way, the peculiar reaction-diffusion dynamics taking place
on the surface of the cell membrane works as a powerful amplifier of slight
anisotropies in the distribution of the external chemical signal.

In this statistical mechanical point of view,
random and gradient-driven polarization appear
as two faces of the same coin,
in good agreement with some of the 
existing biological intuition {\cite{WL03}}.

To better understand the process of spontaneous and gradient-driven cell
polarization from a physical point of view it is convenient to describe the
corresponding signaling network in abstract terms, {\tmem{i.e.}} forgetting
about the particular nature of the molecules involved and considering only the
general structure of the network. This approach has the potential to provide a
unified description of polarization phenomena in distant biological systems.

\begin{figure}[ht]
\centering
  \resizebox{5cm}{!}{\includegraphics{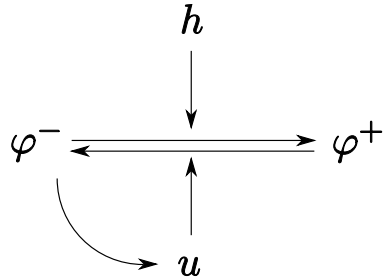}}

  \caption{\label{fig:one}Local structure of a prototypical signaling network
  for cell polarization. Here $\varphi^+$ and $\varphi^-$ represent the local
  concentration of distinct signaling molecules, or distinct states of the
  same molecule, which are converted into each other by the couple of
  counteracting enzymes $h$ and $u$. The $u$ enzymes are activated by
  $\varphi^-$, resulting in an amplification loop. The surface distribution of
  $h$ enzymes is assumed to simply mirror an external distribution of soluble
  chemical attractant. The signaling molecules $\varphi^+$, $\varphi^-$ are
  permanently bound to the cell membrane and perform diffusive motions on it,
  while the $u$ enzymes are free to shuttle between the cytosolic reservoir
  and the membrane.}
\end{figure}

In our abstract signaling network (Fig. \ref{fig:one}) a system or receptors
transduces an external distribution of chemical attractant into an internal
distribution of activated enzymes $h$, which catalyze the switch of a
signaling molecule between two states, that we denote here as $\varphi^-$ and
$\varphi^+$. A counteracting enzyme $u$ transforms the $\varphi^+$ state back
into $\varphi^-$. The molecule $\varphi^-$ in turn activates $u$, thus
realizing a positive feedback loop.

The signaling molecules $\varphi^+$, $\varphi^-$ are permanently bound to the
cell surface $S$ and perform diffusive motions on it, while the $u$ enzymes
are free to shuttle between the cytosolic reservoir and the membrane. In a
more complete description we should consider that also the $h$ enzymes are
shuttling from the cytosol to the membrane {\cite{GCT+05,GKL+07}}. Here
however for simplicity we represent with $h$ only the receptor-bound
fraction{\hspace{1pt}}, which we identify with the external activation field. 
The diffusivity of $u$
enzymes in the cytosol is much higher than the diffusivity of $\varphi^+,
\varphi^-$ molecules on the cell membrane, therefore membrane-bound $u$
enzymes may be assumed to be in approximate equilibrium with the $\varphi^+,
\varphi^-$ concentration field. This fact leaves only the $\varphi^+,
\varphi^-$ surface molecule concentration as relevant dynamic variables.
Moreover, since the $\varphi^+, \varphi^-$ molecules may only be converted
into each other, we are left with only one relevant degree of freedom, their
difference $\varphi \equiv \varphi^+ - \varphi^-$.

The model of Fig. \ref{fig:one} was initially introduced to describe
chemotactic polarization in higher eukaryotes {\cite{GCT+05,GKL+07}}. In that
case, we identify $\varphi^-$ and $\varphi^+$ with PIP2 and PIP3, $u$ with
activated PTEN, and $h$ with activated PI3K.

Recently, it has been proposed that polarization of budding yeast (a lower
eukaryote) may be the result of an amplifying feedback loop similar to the one
described in Fig. \ref{fig:one} {\cite{AAW+08}}. In our language $\varphi^-$
and $\varphi^+$ represent there the activated and unactivated states of the
Cdc42 {\tmem{small GTPase}}{\hspace{1pt}} (see \ref{app:lattice}),
while $u$ would be identified with the activating factor Cdc24. The model of
Ref. {\cite{AAW+08}} lacks a counteracting enzyme playing the role of $h$ in
the scheme of Fig. \ref{fig:one}, and is therefore not bistable. For this
reason, it can reproduce only stochastic, intermittent polarization, as is
observed at the border of the bistability region in the case of chemotactic
polarization {\cite{GCT+05,PRG+04}}. However, in a recent work {\cite{TGH+07}}
a counteracting Cdc42 deactivating factor that could play the role of $h$ has
been described. This suggests that polarization of budding yeast cells may be
driven by a bistable potential allowing the realization of stable
polarization, similarly to the case of chemotactic polarization of higher
eukaryotes.

\section{\label{sec:macroscopic}Macroscopic description of cell polarization}

Cell polarization is a macroscopic effect, emerging from the stochastic
dynamics of a network of chemical reactions taking place in occasion of the
random encounters of specific signaling molecules which perform diffusive
motions and shuttle between the cell cytosol and membrane
{\cite{RSB+03,LH96,WL03}}. A large amount of information has been collected in
recent years about the biochemical aspects of cell polarization in higher
{\cite{RSB+03,LH96,WL03,PD99,Par04,HKK07,PRG+04}} and lower
{\cite{WAW+03,WWS+04,MWL+07}} eukaryotes. However, available data cannot be
considered yet complete or quantitative to a satisfactory degree. This kind of
situation is typical of present efforts to derive macroscopic aspects of cell
behavior from noisy and yet poorly quantitative data about the relevant
microscopic interactions. It is therefore extremely important that a sensible
macroscopic description of cell polarization can be given, starting only from
the knowledge of a few robust properties of the biophysical system.

In this Section we develop such a description. In the next Section we show how
known examples of cell polarization fit in our general scheme.

Let us start here by assuming that we have knowledge only about the following
robust properties of the cell polarization process:
\begin{enumeratenumeric}
  \item {\tmem{Single component order parameter:}} The state of the system may
  be effectively described in terms of the configurations of a
  single-component concentration field $\varphi$ describing the distribution
  of a set of signaling molecules on the cell membrane~$S$.
  
  \item {\tmem{Bistability:}} The underlying chemical reaction networks allows
  the realization of distinct, locally stable chemical phases.
  
  \item {\tmem{Self-tuning:}} A global feedback mechanism controls the
  metastability degree $\psi$ of the system and drives it towards a state of
  phase coexistence.
  
  \item {\tmem{Non-conserved field:}} There are no local constraints on the
  values assumed by the field $\varphi$.
\end{enumeratenumeric}
The present set of properties stems from the abstraction of known properties
of eukaryotic polarization (see also the next Section). In particular,
Property 4. is the consequence of the fast diffusion of $u$ enzymes across the
cytosolic volume {\cite{GKL+07}}. (It is worth mentioning here that our
framework would still hold, although with a few differences, also in the case
that Property 4. be substituted by a local conservation condition.)

{\tmem{Property 1.}} implies that the evolution of the state of the system can
be described by a single stochastic reaction-diffusion equation. Studies of
non-equilibrium statistical mechanics have shown that a few classes of
nonlinear stochastic equations may emerge from the coarse-graining of
microscopic dissipative dynamical systems, depending on general properties,
such as the number of field components and the presence, or absence, of local
conservation laws {\cite{HH77,Bra94}}.

{\tmem{Property 4.}} leads us to select the {\tmem{time-dependent
Landau-Ginzburg model}}
\begin{eqnarray}
  \partial_t \varphi (\mathbf{r}, t) & = & - \frac{\delta \mathcal{F}_{\psi,
  h} [\varphi]}{\delta \varphi (\mathbf{r}, t)} + \Xi (\mathbf{r}, t) 
  \label{langinz}
\end{eqnarray}
(or {\tmem{model A}} in the classification of Ref. {\cite{HH77}}) where
\begin{eqnarray}
  \mathcal{F}_{\psi, h} [\varphi] & = & \int_S \left[ \frac{D}{2}  \left|
  \nabla \varphi \right|^2 + V_{\psi, h} (\varphi) \right] \bignone \mathd
  \mathbf{r}  \label{eq:freeen}
\end{eqnarray}
is an effective free energy functional, $h$ is an external activation field,
$D$ is a diffusion constant, $V_{\psi, h}$ is an effective potential, and
$\Xi$ is a noise term taking into account the effect of thermal agitation and
chemical reaction noise.

{\tmem{Property 2.}} implies that the effective potential $V_{\psi, h}$ has
two potential wells, corresponding to a couple of distinct, stable chemical
phases $\varphi_+$ and $\varphi_-${\hspace{1pt}}{\footnote{We are using a
slightly different notation to distinguish the values $\varphi_+$, $\varphi_-$
assumed by the $\varphi$ field from the names of the concentration fields
$\varphi^+$, $\varphi^-$ of signaling molecules.}}. The kinetic advantage of
transforming a region of $\varphi_+$ phase into a region of $\varphi_-$ phase
is measured by the metastability degree
\begin{eqnarray*}
  \psi & = & V_{\psi, h} (\varphi_+) - V_{\psi, h} (\varphi_-)
\end{eqnarray*}
The polarized state corresponds to the stable coexistence of the $\varphi_+$
and $\varphi_-$ phases in complementary regions of the cell membrane.

{\tmem{Property 3.}} implies that $\psi$ is an integral functional of the
field configuration, going to zero for large times under stationary
conditions. A reasonable analiticity assumption then leads to the following
system of equations, describing the dynamics of cell polarization in the
presence of a stationary external activation field:
\begin{eqnarray}
  \partial_t \varphi (\mathbf{r}, t) & = & D \nabla^2 \varphi (\mathbf{r},
  t) - \frac{\partial V_{\psi, h}}{\partial \varphi} \left[ \varphi
  (\mathbf{r}, t) \right] + \Xi (\mathbf{r}, t)  \label{eq:evo}\\
  \psi (t) & \propto & \int_S \varphi (\mathbf{r}, t) \mathd \mathbf{r}-
  \int_S \varphi (\mathbf{r}, \infty) \mathd \mathbf{r}, \hspace{2em} t
  \rightarrow \infty  \label{eq:constr}
\end{eqnarray}

\section{Model free energy}

It is possible to derive a concrete realization of the scheme described in the
previous Section in the case of the signaling network of Fig. \ref{fig:one} by
using the law of mass action, the quasistationary approximation for enzymatic
kinetics, and the limit of fast cytosolic diffusion 
(see \ref{app:one}). In this case, the state of the system can be described by the
single concentration field $\varphi = \varphi^+ - \varphi^-$, thus giving
{\tmem{Property 1}} of the previous Section. The $\varphi$ field is not
constrained by a local conservation law because $\varphi^+$ molecules can be
freely converted into $\varphi^-$ molecules and back on any point of the cell
surface. This fact corresponds to {\tmem{Property 4.}}

The evolution of the $\varphi$ field is described by the equation
\begin{eqnarray*}
  \partial_t \varphi & = & D \nabla^2 \varphi - k_{\mathrm{cat}} K_\mathrm{ass} f
  \frac{c^2 - \varphi^2}{2 K + c + \varphi} + 2 k_\mathrm{cat} h \frac{c - \varphi}{2
  K + c - \varphi} + \Xi
\end{eqnarray*}
where $f = u_{\mathrm{free}}$ is the volume concentration of free cytosolic $u$
enzymes (which is approximately uniform as a consequence of fast cytosolic
diffusion), $h$ is a surface activation field, $K_{\mathrm{ass}}$ is the
association constant of $u$ enzymes to $\varphi^-$ signaling molecules,
$k_{\mathrm{cat}}$ is a catalytic rate, $K$ is a saturation (Michaelis-Menten)
constant.

The corresponding effective potential has the form $V_{f, h} (\varphi) = fV_1
(\varphi) + hV_2 (\varphi)$ (see \ref{app:one}). The metastability
degree $\psi$ is therefore a function of $h$ and $f$. If $h = h (\mathbf{r},
t)$ is not uniform ({\tmem{e.g.}} if the cell is exposed to a chemical
activation gradient) $\psi$ takes on different values in different points of
the membrane surface. We consider however for the moment the simplest case
where the activation field is uniform in space and constant in time.

A simple analysis shows (\ref{app:one}) that there are regions of
parameter values such that $V_{\psi, h}$ is bistable, with two potential wells
$\varphi_+$ and $\varphi_-$ corresponding to stable phases respectively rich
in the $\varphi^+$ and $\varphi^-$ signaling molecules. Thus {\tmem{Property
2}} is verified.

\begin{figure}[ht]
\centering
  \resizebox{7cm}{!}{\includegraphics{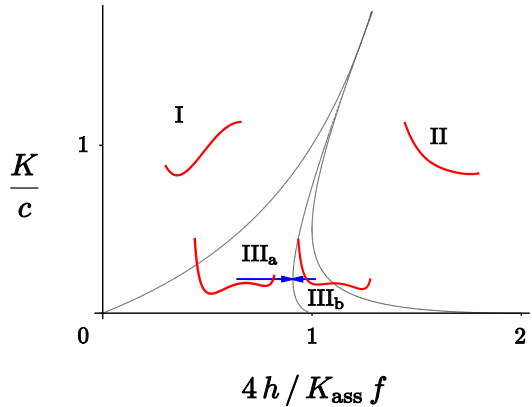}}
  \caption{\label{fig:two}Typical form of the effective potential $V_{\psi, h}
  (\varphi)$ (red curves) in different parameter regions ({\tmem{cf.}}
  supplementary text of Ref. {\cite{GCT+05}}). Region I: one equilibrium point
  $\varphi_- < c$; Region II: one equilibrium point $\varphi_+ = c$; Region
  III$_{\mathrm{a}}$: one stable equilibrium $\varphi_- < c$ and one
  metastable equilibrium $\varphi_+ = c$. Region III$_{\mathrm{b}}$: same as
  III$_{\mathrm{a}}$, but now $\varphi_+ = c$ is stable and $\varphi_-$ is
  metastable. The phase coexistence curve separating III$_a$ from III$_b$ is
  defined by the condition $V_{\psi, h} (\varphi_-) = V_{\psi, h}
  (\varphi_+)$. Arrows show the direction of the dynamic drift towards the
  phase coexistence curve.}
\end{figure}

In the present problem, the volume concentration $f$ of free enzymes varies
in time (but not in space). More information about its values can be gotten
in the limit (realized for small membrane diffusivities and large times) when
the interface between the $\varphi_+$ and the $\varphi_-$ phase is much
smaller than the typical domain size, allowing us to use the so-called
{\tmem{thin wall approximation}}. Then, the value of $f$ is simply linked (see
\ref{app:one}) to the area covered by the $\varphi_-$ phase:
\begin{eqnarray*}
  f (t) - f (\infty) & \propto & \int_S \varphi (\mathbf{r}, t) \mathd
  \mathbf{r}- \bignone \int_S \varphi (\mathbf{r}, \infty) \mathd
  \mathbf{r} \propto \psi (t)
\end{eqnarray*}
showing that {\tmem{the metastability degree is proportional to the excess
fraction of free cytosolic $u$ enzymes}} with respect to their value at
equilibrium. The presence of this global feedback mechanism corresponds to
{\tmem{Property 3.}}

The present situation is closely reminiscent of the decay of the uniform,
metastable state of a {\tmem{supersaturated solution}} with the formation of
precipitate grains {\cite{LP81}}. In that case, the metastability degree is
proportional to the excess solute concentration with respect to its
equilibrium value. The main difference with the present case is that in the
case of precipitation, the density field $\varphi$ is locally constrained by
the law of particle conservation {\cite{Bra94}}, and its evolution is
described by Model B of Ref. {\cite{HH77}}, instead than by model A.

\section{\label{sec:kin}Phase separation kinetics}

In polarization experiments cells are exposed to uniform or gradient
distributions of attractant factors and polarize either spontaneously, or in
the direction of the attractant gradients {\cite{WL03}}. The properties of the
model free energy described in the previous Section and numerical simulations
of a model system {\cite{GCT+05}} suggest that the introduction of an external
attractant distribution moves the system in a region of bistability, where the
uniform phase realized at initial time becomes metastable and germs of a new
phase are nucleated. Depending on the way the system is prepared at initial
time, the metastable phase can be either a $\varphi^+$ rich or a $\varphi^-$
rich phase.

\begin{figure}[ht]
\centering
  \resizebox{10cm}{!}{\includegraphics{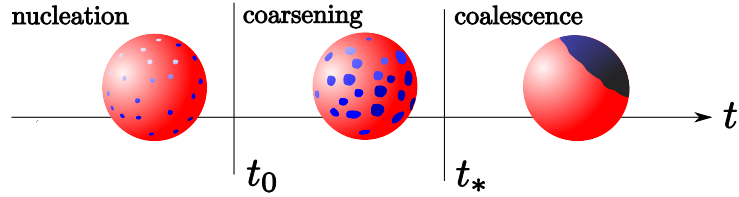}}
  \caption{\label{fig:three}Stages of polarization kinetics.}
\end{figure}

The process of decay of a metastable state in physical systems described by
systems of equations similar to (\ref{eq:evo}, \ref{eq:constr}) has been
extensively studied in the framework of the theory of first-order phase
transitions {\cite{LP81,Bra94}}. The process passes through successive stages
of nucleation, coarsening, and coalescence (Fig. \ref{fig:three}). In the
first stage, approximately circular germs of the new, stabler phase are
produced in the sea of the metastable phase by random fluctuations, or by the
presence of nucleation centers. In the second stage, a process of coarsening
is observed, where larger domains of the new phase grow at the expense of
smaller ones, the average size of domains grow, and the average number of
domains decreases. In a finite system, the process is concluded when a state
of phase coexistence is reached. In this final state, the two phases are in
equilibrium and are polarized in two large complementary domains.

For our purposes, a detailed knowledge of the initial, nucleation
stage{\hspace{1pt}}{\footnote{And therefore of the precise characteristics of
the noise term $\Xi$ which is its driving force.}} is not necessary, as long
as its characteristic time $t_0$ is so fast that a large number of germs of
the new phase is nucleated all over the cell surface, well before the
coarsening stage starts{\hspace{1pt}}{\footnote{The converse case, where $t_0$
is the largest timescale of the problem and polarization is the result of the
rare nucleation of a solitary domain, cannot provide a mechanism of gradient
sensing which is at the same time insensitive to the uniform component of the
attractant field, and highly sensitive to its gradient component. Indeed, the
nucleation of a single domain could provide a mechanism of gradient sensing
only if the gradient would induce significantly different domain nucleation
rates in different points of the cell membrane. But in that case, also
variations in the uniform component of the attractant field would produce
large variations in the typical polarization times, while the converse has
been reported.}}.

To understand the subsequent, coarsening state we have to focus on the laws by
which the domains of the new phase either grow or shrink.

We consider here the case when the new phase is a minority phase, so that we
can restrict our consideration to approximately circular domains, which are
dominating because they minimize the linear tension between the two phases.
For simplicity, we shall also restrict to domains which are small enough that
membrane curvature may be neglected.

\begin{figure}[ht]
\centering
  \begin{tabular}{llllll}
    {\textbf{a}} & \resizebox{6.5cm}{!}{\includegraphics{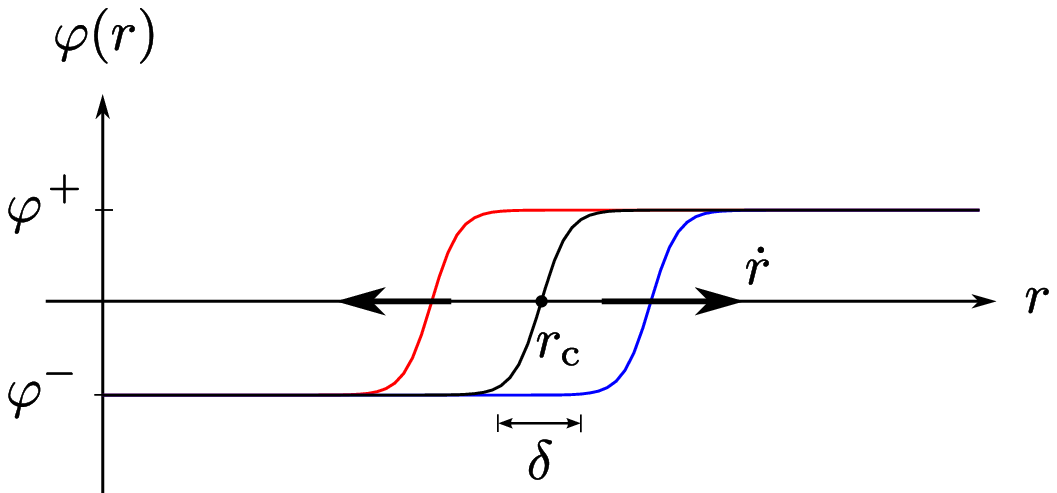}}
     & {\textbf{b}} &
    \resizebox{3.9cm}{!}{\includegraphics{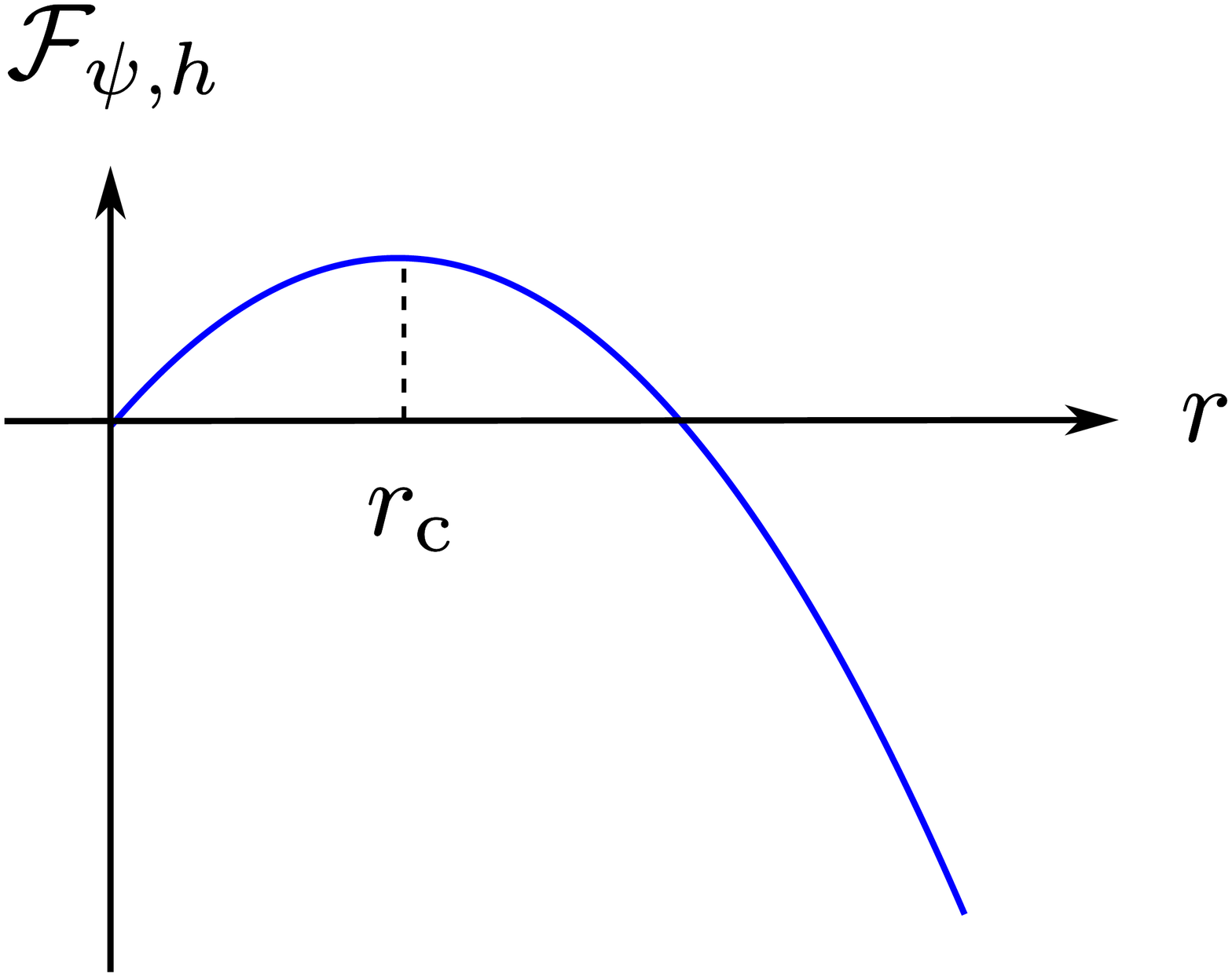}}
  \end{tabular}
  \caption{\label{fig:quattro}{\textbf{a:}} Radial profile of a growing ($r
  > r_c$) and a shrinking ($r < r_c$) circular domain of the $\varphi_-$ phase
  in the sea of the $\varphi_+$ phase. The $\varphi_+$ and the $\varphi_-$
  phase are separated by a diffusive interface of
  thickness{\hspace{3pt}}$\delta$. {\textbf{b:}} Qualitative graph of the
  effective free energy of a circular domain of size $r$. Domains larger than
  $r > r_c$ tend to grow because the energetic gain due to the surface term
  $\pi r^2 \psi$ overcompensate the loss due to a longer interface. The
  converse happens for domains with $r < r_c$.}
\end{figure}

An approximate equation for the growth of a circular domain of size $r$ may be
derived from (\ref{eq:evo}) in the thin wall approximation. Inserting the
approximate propagating solution $\varphi (\mathbf{R}, t) = \phi (R -
r (t))$ (Fig. \ref{fig:quattro}a) for the radial domain profile in
(\ref{eq:evo}) and integrating over $S$ we get
\begin{eqnarray}
  \frac{\partial \mathcal{F}_{\psi, h} [\phi]}{\partial r} & = & -
  \frac{\partial Q}{\partial \dot{r}} + \xi'  \label{eq:diss}
\end{eqnarray}
where
\begin{eqnarray*}
  Q & = & \frac{\dot{r}^2}{2} \int_S (\phi')^2 \mathd \mathbf{R}
\end{eqnarray*}
is a dissipation function {\cite{LL80}} and $\xi'$ is a noise term.

For a circular domain of radius $r$, $Q \simeq \gamma \pi r \dot{r}^2 $,
where
\begin{eqnarray*}
  \gamma & = & \int_0^{\infty} (\phi')^2 \mathd R =
  \int_{\varphi_-}^{\varphi_+} \sqrt{2 V_{\psi, h} (\phi) / D} \mathd \phi
  \bignone
\end{eqnarray*}
is a kinetic coefficient {\cite{Bra94}}. On the other hand the effective free
energy for a circular domain or radius $r$ is{\hspace{4pt}}{\cite{Bra94}}:
\begin{eqnarray}
  \mathcal{F}_{\psi, h} & = & 2 \pi \sigma r - \pi r^2 \psi 
  \label{eq:freeencirc}
\end{eqnarray}
where $\sigma = D \gamma$ is a linear tension.

From (\ref{eq:diss}, \ref{eq:freeencirc}) we get the following approximate
equation for the growth of a circular domain of size $r$:
\begin{eqnarray}
  \gamma \dot{r} & = & \psi - \frac{\sigma}{r} + \xi  \label{eq:domain}
\end{eqnarray}
where $\xi$ is a noise term.

Eq. (\ref{eq:domain}) shows that domains smaller than the critical radius
\begin{eqnarray*}
  r_c & = & \frac{\sigma}{\psi}
\end{eqnarray*}
are mainly dissolved by diffusion, while germs with $r > r_c$ mainly survive
and grow because of the overall gain in free energy (Fig. \ref{fig:quattro}b).

During the nucleation stage the noise term produces a population of germs of
the new phase of size close to
\begin{eqnarray*}
  r_0 & \sim & r_c \sim \delta
\end{eqnarray*}
in a characteristic time $t_0$. For domains with $r > r_c$ the noise term in
(\ref{eq:domain}) may be neglected and domain growth is an almost
deterministic process.

It is interesting to estimate $r_0 \sim \delta$ in terms of observable
parameters. The thickness $\delta$ can be estimated as
\begin{eqnarray*}
  \delta & \sim & \sqrt{D / b}
\end{eqnarray*}
where $b$ is the potential barrier separating the two phases {\cite{Bra94}}.
The height of the potential barrier may in its turn be estimated dimensionally
from (\ref{ilpotenziale}) as $b \sim k_{\mathrm{cat}} hc$, giving
\begin{eqnarray}
  r_0 & \sim & \delta \sim \sqrt{\frac{D}{k_{\mathrm{cat}}}  \frac{c}{h}} 
  \label{eq:stima}
\end{eqnarray}
Using realistic parameter values ($D \sim 1 \mu m^2 / s$, $k_{\mathrm{cat}} \sim
1 s^{- 1}$, $c / h \sim 10$) we get $r_0 \sim 1 \mu m$.

\section{\label{sec:coarsening}The coarsening stage}

When domains of the new phase occupy an appreciable fraction of the membrane
surface $S$ a coarsening stage sets on. Domain growth makes the degree of
metastability $\psi$ decrease and renders further growth of the new phase more
and more difficult. The critical radius $r_c$ grows with time, so that domains
that earlier had size larger than $r_c$ become undercritical and shrink, and
larger domains grow at the expense of smaller ones. In a large system $r_c$
soon becomes the main length scale in the problem, leading to the appearance
of a scaling distribution of domains of size $r$.

The population of coarsening domains of size $r$ can be described in terms of
the size distribution function {\color{black} $n (r, t)$}, such that $n (r, t)
\Delta r$ is the average number of domains with size comprised between $r$ and
$r + \Delta r$, and the total number of domains at time $t$ is given by
\begin{eqnarray*}
  N (t) & = & \int_0^{\infty} n (r, t) \mathd r
\end{eqnarray*}
The time evolution of $n (r, t)$ implied by (\ref{eq:domain}) is described by
a standard Fokker-Planck equation {\cite{Kam07}}. If we restrict our
consideration to supercritical domains we can neglect the diffusive part of
the Fokker-Planck equation since for them the noise term $\xi$ is
negligible.{\hspace{1pt}} This means that the stochastic nature of the problem
enters mainly in the formation of the initial distribution of germ sizes $n
(r, t_0)$, while for $r > r_c$ the time evolution of $n (r, t)$ is dictated by
the deterministic part of (\ref{eq:domain}). Thus, we are left with the
following kinetic equation:

{\color{black} \begin{eqnarray}
  \gamma \frac{\partial n (r, t)}{\partial t} + \frac{\partial}{\partial r}
  \left[ \left( \psi (t) - \frac{\sigma}{r} \right) n (r, t) \right] & = & 0 
  \label{eq:kin}
\end{eqnarray}}

Eq. (\ref{eq:kin}) contains the unknown function $\psi (t)$, and is therefore
not closed. We obtain a closed system by complementing (\ref{eq:kin}) with the
asymptotic law

{\color{black} \begin{eqnarray}
  \psi (t) & \propto & A_{\infty} - \int_0^{\infty} \pi r^2 n (r, t) \mathd
  {\color{black} } r \bignone  \label{eq:area}
\end{eqnarray}}

obtained from (\ref{eq:constr}) in the thin wall approximation. Here
\begin{eqnarray*}
  A_{\infty} & = & \int_0^{\infty} \pi r^2 n (r, \infty) \mathd {\color{black}
  } r \bignone
\end{eqnarray*}
is the area occupied by the new phase at equilibrium.

\begin{figure}[ht]
\centering
  \begin{tabular}{l}
    \\
    \resizebox{7cm}{!}{\includegraphics{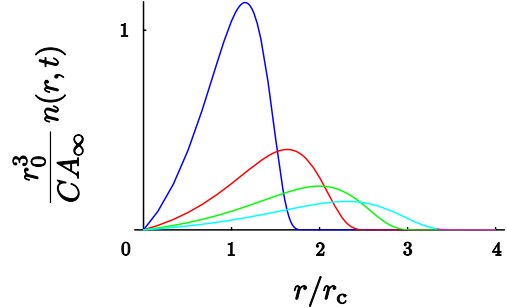}}
  \end{tabular}
  \caption{\label{fig:pdf}Time evolution of the selfsimilar domain size
  distribution $n (r, t)$ ($t / t_0 = 1, 2, 3, 4$).}
\end{figure}

For large times a scaling distribution of domain sizes can be found explicitly
(\ref{app:scaling} and Fig. \ref{fig:pdf}):
\begin{eqnarray}
  n (r, t) \mathd r & = & \frac{CA_{\infty}}{r_c^2} p (r / r_c) \mathd (r /
  r_c), \hspace{2em} \psi (t) = \frac{\sigma}{r_c} \nonumber\\
  r_c & \equiv & r_c (t) = r_0 (t / t_0)^{1 / 2}  \label{eq:loscaling}
\end{eqnarray}
where
\begin{eqnarray}
  p (\rho) & = & {\color{black} \frac{8 \mathe^2 \rho}{(2 - \rho)^4} \exp
  \left( - \frac{4}{2 - \rho} \right)}, \hspace{2em} t_0 = \frac{2 \gamma
  r_0^2}{\sigma},  \label{eq:laprob}
\end{eqnarray}
where $r_0$ is the characteristic domain size at the beginning of the
coarsening stage and $C \simeq 0.11$. 

The total number of domains decreases in time due to the evaporation of small
domains. Using the explicit solution (\ref{eq:loscaling}, \ref{eq:laprob}), we
easily find:
\begin{eqnarray*}
  N (t) = \int_0^{\infty} n (r, t) \mathd r & = & \frac{CA_{\infty}}{r_c^2} =
  \frac{CA_{\infty} / r_0^2}{t}
\end{eqnarray*}
Similarly, it is possible to compute explicitly the value of the average
domain size, which is found to coincide exactly with the critical radius:
\begin{eqnarray*}
  \langle r \rangle & = & r_c
\end{eqnarray*}

\section{\label{sec:spont}Spontaneous and gradient-induced polarization}

The coarsening theory exposed in the previous Section allows to deduce a
simple scaling law for the time needed for spontaneous cell polarization.

If the cell has size $R$, the growth of domains according to
(\ref{eq:loscaling}) comes to a stop at the time $t_{\ast}$ when the average
patch size $\langle r \rangle$ becomes of the order of the cell size $R$. From
(\ref{eq:loscaling}) we get
\begin{eqnarray*}
  t_{\ast} & \sim & t_0  \left( R / r_0 \right)^2
\end{eqnarray*}
At the end of the process the cell is polarized in a random direction. The
actual direction of polarization is the result of the initial random unbalance
in the germ distribution.

The typical time for random polarization is of the order of $10^3\,\mathrm{s}$
{\cite{GKL+07}}. Together with the estimate (\ref{eq:stima}) this gives $t_0
\sim 10\,\mathrm{s}$.

\begin{figure}[ht]
\centering
  \resizebox{4cm}{!}{\includegraphics{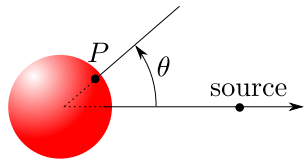}}
  \caption{\label{fig:source}Geometry of stimulation of a cell by an external
  source of attractant.}
\end{figure}

Let us now consider the case where a source of external attractant is present
at some distance from the cell, in such a way that a gradient of external
attractant is created by diffusion close to the cell surface (Fig.
\ref{fig:source}).

The inhomogeneity in the distribution of attractant induces a similarly
inhomogeneous distribution of activated enzymes $h$. This way, the degree of
metastability $\psi$ takes on different values on different points of the cell
surface.

If the cell membrane has a nearly spherical form and a radius $R$ much
smaller than the characteristic scale of the attractant distribution, and if
the gradient component of the activation field is small with respect to the
background component on the scale $R$, the metastability degree $\psi$ at the
beginning of the coarsening process may be written as the sum of a uniform
component $\psi$ and a small space-dependent perturbation:
\begin{eqnarray*}
  &  & \psi + \delta \psi \hspace{2em} \mathrm{with} \hspace{2em} \delta \psi =
  - \epsilon \psi_0 \cos \theta
\end{eqnarray*}
where $\psi_0$ is the value of the uniform component at the beginning of the
coarsening process and $\epsilon$ is the relative gradient on the scale $R$.
The perturbation modifies the equation of domain growth (\ref{eq:domain}) as
follows:
\begin{eqnarray}
  \gamma \dot{r} & = & \psi - \frac{\sigma}{r} - \epsilon \psi_0 \cos \theta +
  \xi  \label{cinque}
\end{eqnarray}
where $\theta$ is an azimuthal angle defined in Fig. \ref{fig:source}.

The uniform component $\psi$ varies in time together with the (approximately)
uniform concentration of $u$ molecules in the cell volume. On the other hand,
the perturbation $\delta \psi$ is constant in time, but not uniform in space,
being proportional to the external attractant distribution.

As long as $\epsilon \psi_0 \ll \psi$, the effect of the perturbation is
negligible, so domain growth proceeds according to the law
(\ref{eq:loscaling}) and the uniform component $\psi$ decays as $t^{- 1 / 2}$.

\begin{figure}[ht]
\centering
  \resizebox{10cm}{!}{\includegraphics{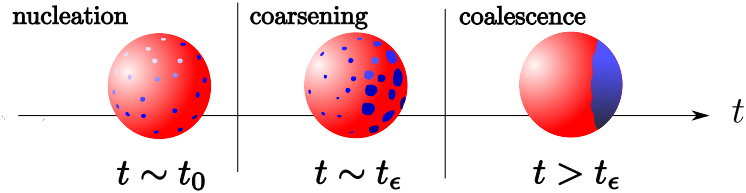}}
  \caption{\label{fig:aniso}The crossover time $t_{\epsilon}$ separates an
  initial stage of isotropic coarsening from a final stage when domains
  evaporate from the back of the cell and condense in the front. }
\end{figure}

In a large cell there is a crossover time $t_{\epsilon}$ when the
perturbation becomes of the same order of the uniform component:
\begin{eqnarray*}
  \psi (t_{\epsilon}) & = & \epsilon \psi_0
\end{eqnarray*}
Using the scaling law (\ref{eq:loscaling}) we get
\begin{eqnarray*}
  t_{\epsilon} & = & \frac{t_0}{\epsilon^2}
\end{eqnarray*}
After $t_{\epsilon}$ domain growth enters in a new stage, where the growth
becomes anisotropic. Domains in the front and back of the cell get different
average sizes (Fig. \ref{fig:aniso}).

Indeed, for $t > t_{\epsilon}$ the leading term in (\ref{cinque}) is the
perturbation $\epsilon \psi_0 \cos \theta$, implying that in the region closer
to the source of the perturbation ($\cos \theta > 0$) the $\varphi_-$ phase
evaporates, and in the region away from the source ($\cos \theta < 0$) it
condenses. At the end of the process, \ complete polarization is realized
(Fig. \ref{fig:aniso}). In this final stage domains grow approximately
linearly in time, thus the total time $t_{\epsilon}'$ to reach polarization is
still a quantity of order $t_{\epsilon}$ (using definition (\ref{smallestgr})
from the next Section it can be estimated as $\frac{1}{2}  \left(1+ \epsilon /
\epsilon_{\mathrm{th}}  \right) t_{\epsilon}$).

The above scheme is valid as soon as the initial nucleation time $t_0$ is
significantly smaller than $t_{\epsilon}$, an assumption which is compatible
with the observation of real {\cite{PRG+04}} and numerical {\cite{GCT+05}}
experiments.

\begin{figure}[ht]
\centering
  \resizebox{7cm}{!}{\includegraphics{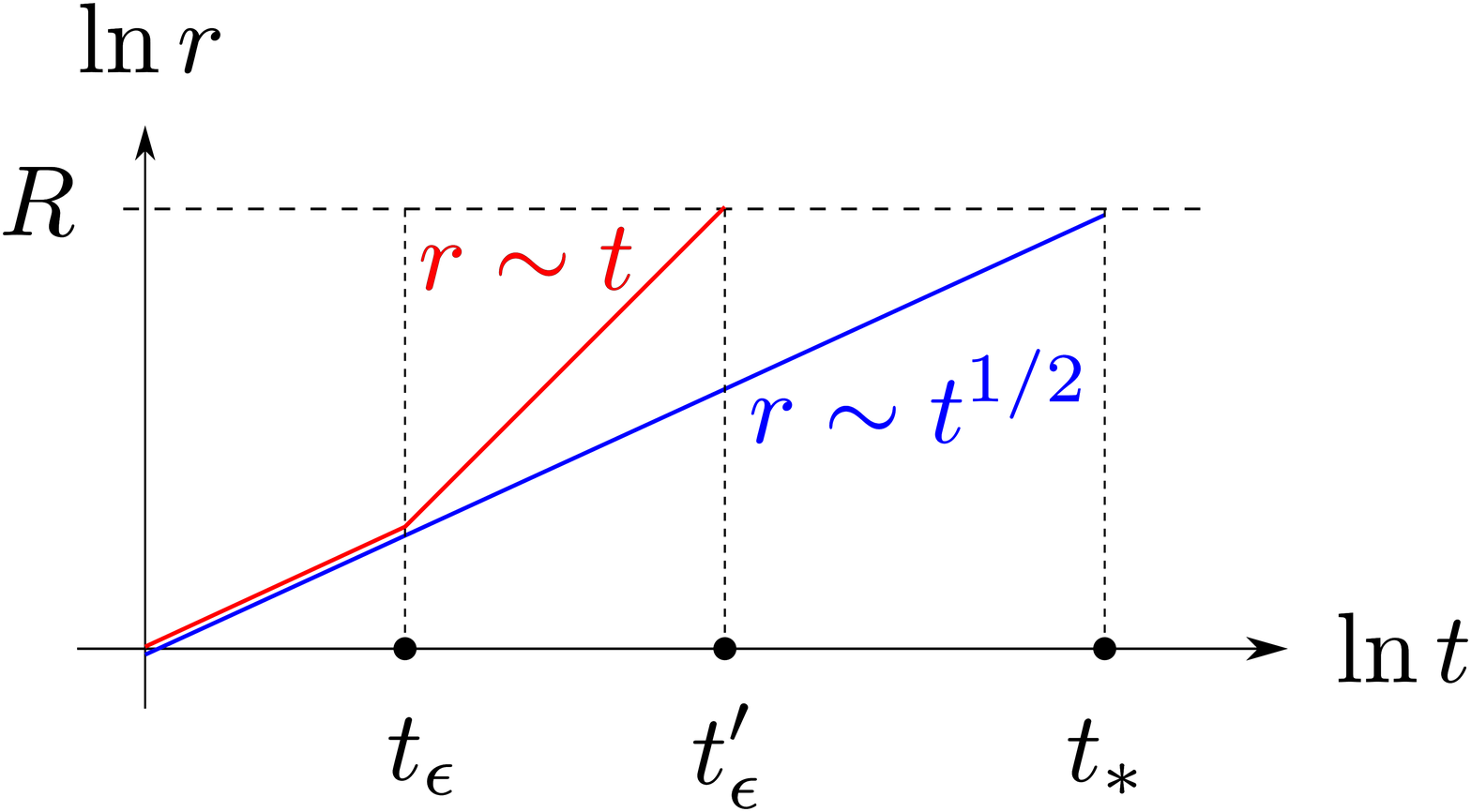}}
  \caption{Dynamic mechanism of gradient sensing. In the presence of a
  gradient (red) domain sizes grow initially as $t^{1 / 2}$. After the
  crossover time $t_{\epsilon}$ they grow linearly in time and the cell
  polarizes in the direction of the gradient (see also Fig. \ref{fig:aniso}).
  In the absence of a gradient (blue) the $t^{1 / 2}$ growth goes on until the
  cell polarizes in a random direction.}
\end{figure}

\section{Gradient sensitivity}

The second stage of domain evolution described in the previous Section occurs
only if $t_{\ast} > t_{\epsilon}$. Otherwise, the presence of a gradient of
attractant becomes irrelevant and only the stage of isotropic domain growth
actually occurs. This condition implies that a {\tmem{{\tmem{smallest
detectable gradient}}}} exists, such that directional sensing is impossible
below it. The threshold value {\color{black} $\epsilon_{\mathrm{th}}$} for
$\epsilon$ is found by the condition{\color{blue}  {\color{black}
$t_{\epsilon} = t_{\ast}$}}. Since the product $\psi r_c$ is a
time-independent constant, we can simply compare its value at initial and
final time when $\epsilon = \epsilon_{\mathrm{th}}$, obtaining that the
{\tmem{threshold detectable gradient}} is
\begin{eqnarray}
  \epsilon_{\mathrm{th}} & = & \frac{r_0}{R}  \label{smallestgr}
\end{eqnarray}
Using the estimates from Sections \ref{sec:kin} and \ref{sec:spont}, and the
typical value $R \sim 10\, \mu \mathrm{m}$, we get $\epsilon \sim 10\%$, a value which is
compatible with the observations {\cite{SNB+06}}.

An interesting speculation is that the bound (\ref{smallestgr}) may explain
why spatial directional sensing was developed only in large eukaryotic cells
and not in smaller prokaryotes, whose directional sensing mechanisms rely
instead on the measurement of temporal variations in concentration gradients
{\cite{ASB+99}}. By solving (\ref{smallestgr}) in terms of the size $R$ we get
the following bound for the size of a cell which may be able to sense a
relative gradient $\epsilon$:
\begin{eqnarray*}
  R & > & \frac{r_0}{\epsilon}
\end{eqnarray*}
Our bound goes in the same direction as the size criterion formulated in
{\cite{BP77}}, but it's independent of it, since the criterion of Ref.
{\cite{BP77}} is based on estimates of signal-to-noise ratios, while our bound
stems from the intrinsic properties of polarization dynamics.

\section{External fluctuations}

One may wonder whether a cell may become polarized by transient gradients
produced by a spontaneous fluctuation in the external distribution of
attractant molecules, or fluctuations in receptor-ligand binding, as has been
suggested in the literature {\cite{LH96}}. Since eukaryotic cells typically
carry 10$^4$--10$^5$ receptors for attractant factors, one expects spontaneous
fluctuations in the fraction of activated receptors to be of the order of
10$^2$, a value which is comparable to observed anisotropy thresholds.
However, to actually produce directed polarization the fluctuation should
sustain itself for several minutes, i.e. for a time comparable to the
characteristic polarization time (such as $t_{\epsilon}$ ). Such an event has
very low probability of being observed since the correlation time of the
fluctuations determined by attractant diffusion at the cell scale and the
characteristic times of receptor-ligand kinetics are much less than the
polarization time. Indeed, the diffusion time is $\sim 1\,$s at the typical cell
size $10\,\mu$m, and the characteristic times of receptor-ligand kinetics are
also $\sim 1\,$s (see online supporting information to Ref. {\cite{SNB+06}}).
Therefore, the direction of cell polarization in the case of a homogeneous
distribution of attractant can only be determined by the inhomogeneity in the
initial distribution of the positions of PIP2-rich germs produced by thermal
fluctuations.

\section{Conclusions}

By using standard statistical mechanical methods we have shown that the
dynamics of signaling domains in cell polarization is independent on the
nature of the signaling molecules and the values of kinetic rate constants, as
long as some very general conditions are met:
\begin{enumeratealpha}
  \item \label{prop:a}Timescale separation allows to describe the polarization
  process in terms of a single concentration field of signaling molecules on
  the cell membrane{\hspace{1pt}}{\footnote{We should consider adding here the
  condition that the concentration field is not locally constrained by a
  conservation law. However, also the converse case of a locally conserved
  field can be treated in a similar way without substantially changing the
  present scheme.}}.
  
  \item \label{prob:b}The underlying chemical reaction network is bistable.
  
  \item \label{prop:c}A global feedback mechanism drives the system towards
  phase coexistence.
  
  \item \label{prop:d}The cell is sufficiently larger than the size of
  nucleating germs of the new phase.
\end{enumeratealpha}
These conditions allow the cell to work as a detector of slight gradients of
external stimulation gradients.

The property of {\tmem{universality}} arising from our analysis cannot be
underestimated. Presently, several efforts are made to understand the
dynamical behavior of living beings starting from microscopic informations
provided by molecular biology. However, these informations are mostly
incomplete and poorly quantitative, and theories that depend in a sensitive
way on them are likely to be of little utility. But if some behavior happens
to be {\tmem{universal}}, a consistent physical theory of it may be built,
which can be compared to experiments.

The universal properties of cell polarization emerge from properties of domain
growth which have been extensively studied in first-order phase
transitions{\hspace{3pt}}{\cite{Bra94}}. The similarity of the two problems
follows from the fact that fast degrees of freedom of chemical kinetics are in
approximate equilibrium with slower degrees of freedom, which can be described
by means of an effective free energy functional. It is worth observing that in
the biological system studied here, there is no direct interaction between
signaling molecules, similar to the one observed in solid state system such as
binary alloys, but only an effective interaction mediated by enzyme activity,
binding, unbinding and diffusion processes.

Our theoretical scheme allows to shed light on some non trivial questions,
such as the mechanism of directional sensing and the effect of random
fluctuations of the medium on the polarization process. Random polarization
appears as the result of the intrinsic stochasticity of the process of domain
nucleation and not of random fluctuations of the medium. Random and
gradient-induced polarization appear as the two sides of a same coin. Our
scheme provides an explanation of why spatial directional sensing is not
observed in the small prokaryotic cells, and provides asymptotic estimates for
polarization times and threshold detectable gradients.

An important component of our picture is the existence of a global coupling of
the degree of metastability to the state of the system {\cite{GKL+07,FCG+08}}.
{\color{black} The constrained phase-ordering dynamics tunes the system
towards phase coexistence}, similarly to what happens in the case of a
precipitating supersaturated solution. The global control allowing self-tuning
to phase-coexistence is realized by shuttling of enzymes from the cytosol to
the cell membrane and backwards.

Some of the features that we have observed in cell polarization have been
considered in previous works, such as the fact that equations of the form
(\ref{langinz}) are relevant for the description of systems of bistable
chemical reactions {\cite{Sch72,Kam07}}, and that global couplings in
activator-inhibitor reaction-diffusion systems may lead to the formation of
stable spatiotemporal patterns {\cite{GM72,Sch00}}. The peculiar properties of
this kind of systems have led to the use of the term of {\tmem{excitable}} or
{\tmem{active}} media. \ Using this same language, we can say that the cell
membrane acts as an active medium responding to the stimulation with the
formation of domains of a new phase. Our work proposes that directional
sensing results from the peculiar, universal features of the phase-ordering
dynamics of these domains.

From a biological point of view, the universality of the polarization process
allows the cell to behave in a robust, predictable way, independent on
microscopic peculiarities such as the precise values of reaction rates and
diffusion constants.

We first proposed that chemotactic cell polarization may result from the
simple ingredients of bistability induced by a positive local feedback loop in
a signaling network and global control induced by shuttling of enzymes between
the cytosol and the membrane in our previous works
{\cite{GCT+05,GKL+07,CGC+07}}. Other authors have proposed similar models,
either independently {\cite{SLN05}} or subsequently {\cite{MXA+06}} (a review
of models of chemotactic polarization can be found in Ref. {\cite{ID07}}).
Some of these models try to take into account computationally the interactions
of a large numbers of chemical factors, while retaining the essential role of
a feedback loop as generator of a phase-separation instability. However, most
of the reaction rates that should be provided to perform such computations are
known with very poor accuracy. Our framework suggests however that such a
detailed description may be not necessary, as long as properties
\ref{prop:a}),...,\ref{prop:d}) are met.

Aspects of the bistable mechanism of eukaryotic polarization firstly
introduced in Ref. {\cite{GCT+05}} (supporting material) have been considered
in recent papers {\cite{BAB08,MJE08}} as relevant to polarization phenomena. A
similar mechanism, out of the bistability region, has been proposed to explain
intermittent polarization in budding yeast {\cite{AAW+08}}. These works
suggest that the combination of bistability and global control
{\cite{GCT+05,GKL+07}} is providing a useful paradigm for the understanding of
cell polarization phenomena.

\paragraph{Acknowledgments}

We thank Guido Serini for many inspiring discussions. This research was
supported in part by the National Science Foundation under Grant No. NSF
PHY05-51164.

\appendix\section{\label{app:lattice}Lattice gas description of cell
polarization}

The signaling molecules PIP2 and PIP3 are different phosphorylation states of
the {\tmem{phosphatidylinositol}} molecule, {\tmem{i.e.}}, they carry a
different number of phosphate groups attached (2 and 3, respectively). Enzymes
which catalyze phosphorylation of their substrate, {\tmem{i.e.}} the addition
of a phosphate group, are called {\tmem{kinases}}, while dephosphorylating
enzymes are called {\tmem{phosphatases}}.

It is natural to visualize the state of a chemical system such as the one
described in Fig. \ref{fig:one} in terms of two families of classical spins on
a twodimensional lattice, taking on values -1 (PIP2, PTEN), 0 (an empty site),
+1 (PIP3, PI3K) {\cite{FCG+08}}. Taking into account fast cytosolic diffusion,
the enzyme family becomes slaved to the substrate family {\cite{FCG+08}}.

In this lattice-gas description the existence of a cytosolic enzymatic
reservoir exchanging enzymes with the cell membrane is represented by a
chemical potential for enzyme creation and destruction (actually, adsorption
and desorption to/from the cell membrane), globally coupled to the lattice
configuration {\cite{FCG+08}}.

The PIP2 and PIP3 molecules constitute approximately 1\% of the total number
of membrane phospholipids, and the number of PI3K and PTEN enzymes are at
least one order of magnitude lower, thus, both the substrate and the enzyme
population should be thought as diluted gases.

Two-state (or multistate) molecules such as PIP2 and PIP3 are all but an
exception in cell biology. Another example is given by {\tmem{small GTPases}},
such as the Cdc42 molecule involved in the polarization of budding yeast,
which can be found either in the activated GTP state or in the deactivated GDP
state. The switch between the two phosphorylation states is catalyzed by a
couple of activating (GEF) and deactivating (GAP) enzymes {\cite{AJL+07}}.

\section{\label{app:one}Mean-field equations for eukaryotic polarization}

We derive here mean-field equations for eukaryotic polarization using standard
methods of chemical kinetics, including Michaelis-Menten saturation terms for
the enzymatic components{\hspace{1pt}}{\footnote{Michaelis-Menten saturation
terms arise from timescale separation in enzymatic kinetics, which allows to
make use of a quasi-stationary approximation{\hspace{3pt}}{\cite{CCT07}}.}}.
We make use of the fact that the diffusivity $D_{\mathrm{vol}}$ of $u$ enzymes
in the cytosol is much faster than the diffusivity $D$ of $\varphi$ molecules
on the cell membrane: this fact allows to considerably reduce the number of
dynamical degrees of freedom.

We describe the macroscopic state of the cell using surface concentration
fields of membrane-bound molecules (Fig. \ref{fig:one}) and the volume
concentration field $f \equiv u_{\mathrm{free}}$ of free $u$ enzymes.

The chemical kinetic equations for the signaling network of eukaryotic
polarization are:
\begin{eqnarray}
  \partial_t \varphi^+ & = & D \nabla^2 \varphi^+ \um k_{\mathrm{c} \mathrm{at}}
  \frac{u \varphi^+}{K + \varphi^+} + k_{\mathrm{\mathrm{cat}}}  \frac{h
  \varphi^-}{K + \varphi^-}  \label{phipiu}\\
  \partial_t \varphi^- & = & D \nabla^2 \varphi_- + k_{\mathrm{c} \mathrm{at}} 
  \frac{u \varphi^+}{K + \varphi^+} - k_{\mathrm{c} \mathrm{at}}  \frac{h
  \varphi^-}{K + \varphi^-} - \partial_t u  \label{phimeno}\\
  \partial_t u & = & k_{\mathrm{a} \mathrm{ss}} f \varphi^- -
  k_{\mathrm{{diss}}} u  \label{psidot}\\
  \partial_t f & = & \nabla \cdot (D_{\mathrm{vol}} \nabla f), \hspace{2em}
  \hspace{2em}  \label{psivol}
\end{eqnarray}
They must be complemented with the boundary condition
\begin{eqnarray}
  J & \equiv & D_{\mathrm{vol}}  \frac{\partial f}{\partial \mathbf{n}}
  = \partial_t u  \label{outflux}
\end{eqnarray}
where $\partial / \partial \mathbf{n}$ is the derivative along the outward
normal to the membrane surface $S$. Condition (\ref{outflux}) expresses the
fact that the flux of $u$ enzymes leaving the cytosolic volume equals the flux
of enzymes being bound to the cell membrane.

For simplicity, we consider here identical catalytic, association and
dissociation rates ($k_{\mathrm{cat}}$, $k_{\mathrm{ass}}$, $k_{\mathrm{diss}}$) and
Michaelis-Menten constants $K$ for the $\varphi^+ \rightarrow \varphi^-$ and
$\varphi^- \rightarrow \varphi^+$ processes. This is compatible with existing
information about these processes, suggesting that reaction rates differ by
factors of order{\hspace{3pt}}1{\hspace{3pt}}{\cite{GCT+05}} and allows to
easily study the equations analytically.

Typical values for surface and cytosolic diffusivity are $D \sim 1 \mu m^2 /
s$, $D_{\mathrm{vol}} \sim 10 \mu m^2 / s${\hspace{3pt}}{\cite{GCT+05}}. Typical
values for rate constants are: $k_{\mathrm{cat}} \sim k_{\mathrm{diss}} \sim 1
s^{- 1}$, $k_{\mathrm{ass}} \sim 0.05 s^{- 1} \mathrm{nM}^{- 1}$; for the total
number of $\varphi^+$ and $\varphi^-$ molecules, and the total number of $u$
and $h$ enzymes: $N_{\varphi} \sim 10^6$, $N_u \sim N_h \sim
10^4$--$10^5$.
Observe that $N_u / N_{\varphi} \ll 1$.

The usual definition of macroscopic fields such as $u$ is as follows. For each
point $\mathbf{r}$ in space we choose a volume $v$ centered in
$\mathbf{r}$, containing $n (v)$ molecules, and we compute concentrations as
$u (\mathbf{r}) = \lim_{v \rightarrow 0} n (v) / v$. This implies that the
number of molecules of the relevant chemical factors is so large that $v$ can
be chosen much smaller than the size of the system, but large enough that the
resulting field $\varphi (\mathbf{r})$ is approximately continuous. This
hypothesis is not always acceptable, since enzymatic molecules are present in
the cell in very small numbers. We shall therefore assume that real
concentrations are described as the sum of an average part $u$, described by
mean field equations of the kind (\ref{phipiu}--\ref{outflux}), and a
fluctuating part $\delta u$ taking into account both the discrete character of
the concentration field and thermal disorder. The fluctuations $\delta u$ due
to random adsorption and desorption processes are at the origin of the noise
term $\Xi$ in (\ref{eq:evo}) (see \ref{app:thermal}).

Since enzyme diffusion in the cytosol is faster than phospholipidic diffusion
on the membrane, during the characteristic times of the dynamics of
membrane-bound factors, $f (\mathbf{r}, t)$ relaxes to the approximately
uniform value
\begin{eqnarray}
  f (t) & = & f_0 - \frac{1}{V} \int_S u_- (\mathbf{r}, t) \mathd
  \mathbf{r}  \label{meanpsi}
\end{eqnarray}
where $f_0 = N_u / V$, while $u$ relaxes to the local equilibrium value
\begin{eqnarray}
  u & = & K_{a \mathrm{ss}} f \varphi^-  \label{equcond}
\end{eqnarray}
where $K_\mathrm{ass} = k_\mathrm{ass} / k_{\mathrm{diss}}$.

On the other hand, by summing (\ref{phipiu}) and (\ref{phimeno}) we get
\begin{eqnarray}
  \partial_t \left( \varphi_+ + \varphi_- \right) & = & D \nabla^2 (\varphi_+
  + \varphi_-) - \partial_t u_-  \label{diffusionesemplice}
\end{eqnarray}
Since $N_u / N_{\varphi} \ll 1$ we neglect the term $\partial_t u$. Then,
(\ref{diffusionesemplice}) shows that the sum $c = \varphi^+ + \varphi^-$
tends to be approximately uniform and constant in time.

By subtracting (\ref{phipiu}) and (\ref{phimeno}) and introducing the
difference concentration field $\varphi = \varphi^+ - \varphi^-$ we get
\begin{eqnarray}
  \partial_t \varphi & = & D \nabla^2 \varphi - k_{\mathrm{cat}}  \frac{2 u (c +
  \varphi)}{2 K + c + \varphi} + k_{\mathrm{cat}}  \frac{2 h (c - \varphi)}{2 K
  + c - \varphi}  \label{sempluno}
\end{eqnarray}
and using the local equilibrium condition (\ref{equcond}) we end up with
\begin{eqnarray}
  \partial_t \varphi & = & D \nabla^2 \varphi - k_{\mathrm{cat}} K_{\mathrm{ass}}
  f \frac{c^2 - \varphi^2}{2 K + c + \varphi} + 2 k_{\mathrm{cat}} h \frac{c -
  \varphi}{2 K + c - \varphi}  \label{endup}
\end{eqnarray}
Only values $- c \leqslant \varphi \leqslant \varphi$ correspond to positive
concentrations and are therefore physical.

From (\ref{endup}), (\ref{psidot}) and (\ref{meanpsi}) we get the following
system:
\begin{eqnarray}
  \partial_t \varphi (\mathbf{r}, t) & = & - \frac{\delta \mathcal{F}_{f, h}
  [\varphi]}{\delta \varphi (\mathbf{r}, t)}  \label{reactdiff}\\
  \dot{f} (t) & = & - V^{- 1} k_{\mathrm{ass}} f (t) \int_S \varphi^-  \bignone
  \mathd \mathbf{r}+ k_{\mathrm{diss}}  \left( f_0 - f (t) \right) 
  \label{betadot}
\end{eqnarray}
where
\begin{eqnarray}
  \mathcal{F}_{f, h} [\varphi] & = & \int_S \bignone \left[ \frac{1}{2} D
  \left| \nabla \varphi \right|^2 + V_{f, h} (\varphi) \right] \mathd
  \mathbf{r}  \label{free}\\
  V_{f, h} (\varphi) & = &  \hspace{1em} 2 k_{\mathrm{cat}} hc \left[ - \phi - 2
  \kappa \ln \left( 2 \kappa + 1 - \phi \right) \right] 
  \label{ilpotenziale}\\
  &  & + \frac{1}{2} k_{\mathrm{cat}} K_{\mathrm{ass}} fc^2 [- \phi^2 / 2 +
  \left( 2 \kappa + 1 \right) \phi \nonumber\\
  &  & \left. - 4 \kappa (\kappa + 1) \ln \left( 2 \kappa + 1 + \phi \right)
  \right] \nonumber
\end{eqnarray}
and we make use of the nondimensional variables $\phi = \varphi / c$, $\kappa
= K / c$.

The quantity {\color{black} $\mathcal{F}_{f, h}$ plays the role of a
generalized free energy for the system}, and can be used to study its
approximate equilibria as long as the characteristic times of variation of $f$
are longer than the characteristic times of variation of the $\varphi$ field.

We are interested in parameter values such that (\ref{ilpotenziale}) is
bistable. In what follows we consider the case of constant and uniform
activation field $h$, and constant $f$.

\begin{figure}[ht]
\centering
  \resizebox{9.5cm}{!}{\includegraphics{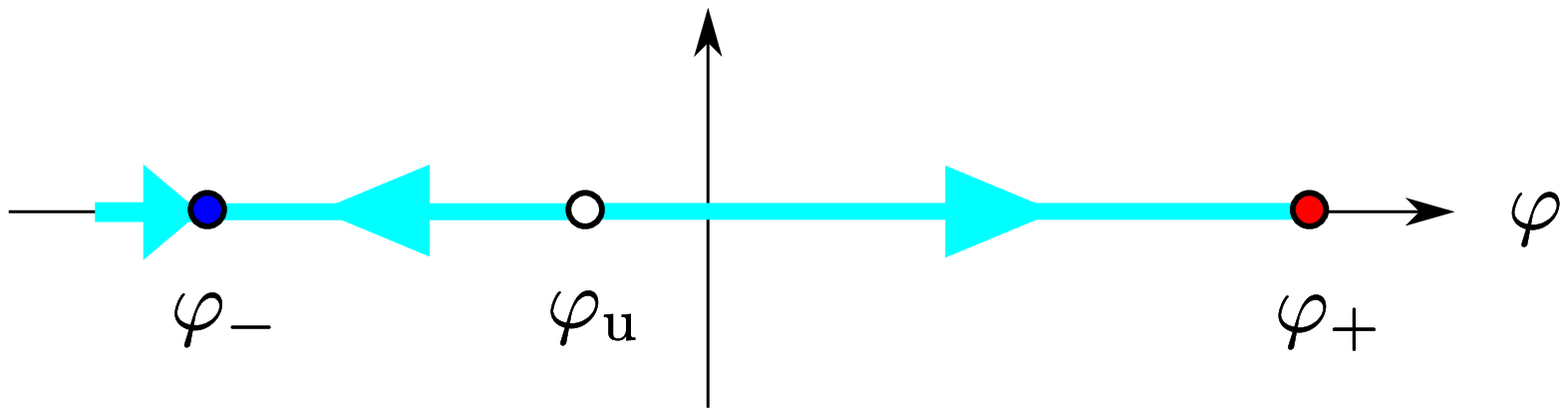}}
  \caption{\label{fig:dyn}Eq. (\ref{endup}) describes a local flow towards
  either a $\varphi^-$ rich or a $\varphi^+$ rich stable phase.}
\end{figure}

The critical points of the effective potential $V_{f, h}$ are
\begin{eqnarray*}
  \phi_- & = & \kappa - \lambda / 2 - \sqrt{\left( \kappa - \lambda / 2
  \right)^2 - (\lambda - 1) (2 \kappa + 1)}\\
  \phi_u  & = & \kappa - \lambda / 2 + \sqrt{\left( \kappa - \lambda / 2
  \right)^2 - (\lambda - 1) (2 \kappa + 1)}\\
  \phi_+ & = & 1
\end{eqnarray*}
where
\begin{eqnarray*}
  \lambda & = & \frac{4 h}{K_{a \mathrm{ss}} f}
\end{eqnarray*}
The potential $V_{f, h}$ is bistable when the three critical points are all
real and physical. In that case, (\ref{endup}) describes a dynamical system
that may locally favor either a $\varphi^-$ rich or a $\varphi^+$ rich stable
phase (Fig. \ref{fig:dyn}).

The two roots $\phi_- < \phi_u$ are real if
\begin{eqnarray}
  \lambda & < & 2 (3 \kappa + 1) - 4 \sqrt{\mathcal{\kappa} (2
  \mathcal{\kappa} + 1)}, \hspace{1em} \lambda > 2 (3 \mathcal{\kappa} + 1) +
  4 \sqrt{\mathcal{\kappa} (2 \mathcal{\kappa} + 1)}  \label{radicireali}
\end{eqnarray}
The l.h.s. condition defines the right boundary of the bistability region of
parameter space (Region III of Fig. \ref{fig:two}).

The two roots are physical ($- 1 \leqslant \phi_- < \phi_u \leqslant 1$) when
\begin{eqnarray}
  \kappa & \leqslant & \frac{\lambda}{2 - \lambda} \hspace{1em} \mathrm{and}
  \hspace{1em} \kappa \leqslant 1 + \frac{\lambda}{2}  \label{radicifisiche}
\end{eqnarray}
The l.h.s. condition defines the left boundary of the bistability region
(Region III in Fig. \ref{fig:two}). The inequality $\phi_- \geqslant - 1$ on
the other hand is always verified if $\lambda < 2$.

The left and right boundaries of Region III meet in the triple point
\begin{eqnarray*}
  \lambda & = & 1 - \sqrt{5}, \hspace{2em} \kappa = (1 + \sqrt{5}) / 2
\end{eqnarray*}
So, the $\lambda$--$\kappa$ plane can be divided into three regions (Fig.
\ref{fig:two} and supplementary text of Ref. {\cite{GCT+05}}). In Region III,
the system has to stable minima $\varphi_+$ and $\varphi_-$, separated by the
unstable equilibrium $\varphi_u$. Outside Region III the potential has a
single minimum, either rich in $\varphi^-$ (Region I) or rich in $\varphi^+$
(Region II).

Region III may be divided in two parts, depending on which phase is stabler.
In Region III$_a$ (Fig. \ref{fig:two}) the stabler phase is $\varphi_-$, while
in Region III$_b$ it is $\varphi_+$. The two subregions are separated by the
phase-coexistence curve $\psi \equiv V_{f, h} (\varphi_+) - V_{f, h}
(\varphi_-) = 0$, where the two stable equilibria $\varphi_+$ and $\varphi_-$
have the same energy.

Close to the phase coexistence curve $\psi$ is much smaller than the potential
barrier separating the two minima. In this region
\begin{eqnarray}
  \psi & \simeq & 2 k_{\mathrm{cat}} hc \left[ \phi_+ - \phi_- + 2 \kappa \ln
  \left( 1 + \frac{\phi_+ - \phi_-}{2 \kappa} \right) \right]  \left(
  \frac{f}{f_{\infty}} - 1 \right)  \label{deltavapprox}
\end{eqnarray}
where the factor $f / f_{\infty} - 1$ represents the excess fraction of free
$u$ enzymes at a given time, with respect to the equilibrium value.

Observe that an actual excess of free enzymes renders the $\varphi_-$ phase
more stable, while a negative excess (a deficit) stabilizes the $\varphi_+$
phase.

If the $\varphi_-$ phase is the more stable one, it tends to occupy larger and
larger regions of the cell surface, thus decreasing
$f${\hspace{3pt}}({\tmem{cf.}} the quasi-equilibrium conditions
(\ref{meanpsi}) and (\ref{equcond})) and its own stability relative to the
$\varphi_+$ phase.

A symmetric situation is encountered if $\varphi_+$ is the more stable phase
at initial time.

Thus, the process of growth of any of the two phases decreases the
metastability degree $\psi$ and drives the system towards a condition of
phase-coexistence ({\tmem{i.e.}} towards a polarized state).

We may wonder whether uniform equilibrium states also exist, that may compete
with polarized states. Looking for stable uniform equilibria $\varphi =
\varphi_-$ in Region III$_a$ gives the algebraic conditions
\begin{eqnarray}
  \lambda & = & \frac{- \phi^2 + 2 \kappa \phi + (2 \kappa + 1)}{\phi + (2
  \kappa + 1)} = 2 \frac{N_h}{N_u}  \left[ \left( 1 + \frac{2
  K_{\mathrm{ass}}}{VN_{\varphi}} \right) - \phi \right]  \label{cond1}\\
  \varphi & \leqslant & 2 \sqrt{\kappa (2 \kappa + 1)} - (2 \kappa + 1) 
  \label{cond3}
\end{eqnarray}
which may be studied graphically, showing that uniform equilibria are
impossible in a large part of Region III, and in particular if
\begin{eqnarray}
  \kappa & < & \frac{1}{2} \hspace{1em} \mathrm{and} \hspace{1em} 2
  \frac{N_h}{N_u}  \left( 1 + \frac{2 K_{\mathrm{ass}}}{VN_{\varphi}} \right) >
  1  \label{ruleout}
\end{eqnarray}
Uniform equilibria do not exist in this region because the total number of $u$
enzymes is not large enough to stabilize a uniform $\varphi_-$ phase extended
along the whole membrane surface.

Instead, uniform equilibria with $\varphi = \varphi_+$ exist, and correspond
to configurations where all $u$ enzymes are free.

\section{Thermal and chemical noise}
\label{app:thermal}
Up to this point we have neglected fluctuations in the number of
membrane-bound enzymes, so that every local minimum of $V_{f, h}$ corresponds
to a stable phase having an infinite lifetime. However, since the number of
bound enzymes molecules in the real system fluctuates locally, the field
$\varphi (\mathbf{r}, t)$ should be seen as a stochastic field.

The fluctuations $\delta f$ around the equilibrium enzyme concentration
$f_{\infty}$ in the volume $V$ due to membrane adsorption and desorption
processes induce fluctuations $\delta u$ around the local equilibrium value
(\ref{equcond}) in the{\color{black} } concentration of membrane-bound
enzymes.

To derive quantitative relations we have to compute the encounter rates of a
free $u$ particle fluctuating in the volume $V$ and a $\varphi_-$ binding site
on the surface $S$.

The adsorption-desorption process can be described by a simple master
equation{\hspace{3pt}}{\cite{Gar83}}. Let us consider that a reservoir of
volume $V$ contains a number $N^{\mathrm{free}} \leqslant N^{\mathrm{tot}}$ of
molecules, which can be adsorbed and desorbed by a small surface element
$\Sigma$ containing $N^\mathrm{b.s.}$ binding sites. One has the mean-field kinetic
equation
\begin{eqnarray*}
  \frac{\mathd}{\mathd t} N^{\mathrm{bound}} & = & k_{\mathrm{ass}} V^{- 1}
  N^\mathrm{b.s.} N^{\mathrm{free}} - k_\mathrm{diss} N^{\mathrm{bound}}
\end{eqnarray*}
which at equilibrium gives
\begin{eqnarray*}
  N^{\mathrm{bound}} & = & \alpha N^{\mathrm{free}} = \frac{\alpha}{1 + \alpha}
  N^{\mathrm{tot}} \hspace{2em} \left( \alpha = K_{\mathrm{ass}} V^{- 1}
  N^\mathrm{b.s.} \right)
\end{eqnarray*}
Let $P_N$ be the probability to observe $N^{\mathrm{bound}} = N$, and
$r^{\pm}_N$ the time rates of the processes $N \rightarrow N \pm 1$. Then the
process is described by the master equation
\begin{eqnarray*}
  \dot{P_N} & = & r^+_{N - 1} P_{N - 1} - (r^+_N + r_N^-) P_N + r^-_{N + 1}
  P_{N + 1}
\end{eqnarray*}
which has the stationary solution
\begin{eqnarray*}
  P_N & = & \prod_{j = 0}^{N - 1} \frac{r^+_j}{r^-_{j + 1}} P_0
\end{eqnarray*}
where $P_0$ is a normalizing factor. Letting
\begin{eqnarray*}
  r^+_N & = & c\,k_\mathrm{ass} (N^{\mathrm{tot}} - N), \hspace{2em} 
  r^-_N = k_\mathrm{diss} N
\end{eqnarray*}
one finds a binomial distribution with
\begin{eqnarray*}
  \langle N^{\mathrm{bound}} \rangle & = & {\color{black} \frac{\alpha
  N^{\mathrm{tot}}}{1 + \alpha} } = \alpha N^{\mathrm{free}}\\
  \langle \left( N^{\mathrm{bound}} \right)^2 \rangle - \langle N^{\mathrm{bound}}
  \rangle^2 & = & {\color{black} \frac{\alpha N^{\mathrm{tot}}}{\left( 1 +
  \alpha \right)^2} } = \frac{N^{\mathrm{bound}}
  N^{\mathrm{free}}}{N^{\mathrm{tot}}}
\end{eqnarray*}
By identifying $f = N^{\mathrm{free}} / V$ in (\ref{equcond}) we can model the
adsorption-desorption noise with a Gaussian noise term $\Xi$ with zero mean
and the correct variance:
\begin{eqnarray*}
  \langle \Xi (\mathbf{r}, t) \Xi (\mathbf{r}', t') \rangle & = & 2 \Gamma
  \delta (\mathbf{r}-\mathbf{r}') \delta (t - t')
\end{eqnarray*}
where
\begin{eqnarray*}
  \Gamma & = & \frac{k_{\mathrm{diss}}}{k_{\mathrm{cat}}}  \frac{\varphi^+}{K +
  \varphi^+}  \frac{f}{f_0} (K_{\mathrm{ass}} f \varphi^-)
\end{eqnarray*}

\section{\label{app:scaling}Scale invariant size distribution}

In the domain coarsening stage described in Section \ref{sec:coarsening}, the
characteristic size $r_c (t)$ of domains grows with time, and soon becomes the
largest scale, so that a scaling distribution of domain sizes arises. In the
asymptotic regime (for large times) it is possible to derive a self similar
solution of the system of equations (\ref{eq:kin}, \ref{eq:area}):

{\color{black} \begin{eqnarray}
  \gamma \frac{\partial n (r, t)}{\partial t} + \frac{\partial}{\partial r}
  \left[ \left( \psi (t) - \frac{\sigma}{r} \right) n (r, t) \right] & = & 0 
  \label{eq:kin1}\\
  \psi (t) \propto A_{\infty} - \int_0^{\infty} \pi r^2 n (r, t) \mathd
  {\color{black} } r \bignone & \rightarrow & 0 \hspace{2em} \mathrm{for}
  \hspace{2em} t \rightarrow \infty  \label{eq:kin2}
\end{eqnarray}}

We start by looking for a solution in the form
\begin{eqnarray}
  n (r, t) & = & \left[ r_c (t) \right]^k g (r / r_c (t))  \label{scalsol}
\end{eqnarray}
It is easy to verify that $k$ must be given the value $- 3$ in order that
(\ref{eq:kin2}) may attain its asymptotic limit.

Substituting (\ref{scalsol}) in (\ref{eq:kin1}), reexpressing the result in
terms of the nondimensional variable
\begin{eqnarray*}
  \rho & = & r / r_c
\end{eqnarray*}
and balancing terms in the resulting equation, we find that an asymptotic
solution for large times may exist only if
\begin{eqnarray*}
  \psi (t) & = & \frac{\sigma}{r_c (t)}, \hspace{2em} r_c (t) = r_0  \left( t
  / t_0 \right)^{1 / 2}
\end{eqnarray*}
and
\begin{eqnarray}
  \left[ {\color{black} - \sigma \rho + \sigma \rho^2 - \frac{1}{2} 
  \frac{\gamma r_0^2}{t_0} \rho^3} \right] g' (\rho) + \left[ {\color{black}
  \sigma - \frac{3}{2}  \frac{\gamma r_0^2}{t_0} \rho^2} \right] g (\rho) & =
  & 0  \label{eq:grho}
\end{eqnarray}
A smooth, positive, normalizable solution of (\ref{eq:grho}) may be found only
when two of the poles of $g' (\rho) / g (\rho)$ coalesce, which gives
\begin{eqnarray}
  t_0 & = & \frac{2 \gamma r_0^2}{\sigma}  \label{eq:separatrice}
\end{eqnarray}
and finally{\hspace{1pt}}{\footnote{We thank Alan Bray for pointing out 
to us that
this problem has been discussed in a different context in Ref.
{\cite{SM95}}.}}
\begin{eqnarray*}
  g (\rho) & = & \left\{ \begin{array}{ll}
    CA_{\infty}  \frac{8 \mathe^2 \rho}{(2 - \rho)^4} \exp \left( - \frac{4}{2
    - \rho} \right) & \mathrm{for} \hspace{1em} 0 \leqslant \rho \leqslant 2\\
    0 & \mathrm{elsewhere}
  \end{array} \right.
\end{eqnarray*}
with
\begin{eqnarray*}
  C & = & \frac{1}{4 \pi [1 + 2 \mathe^2 \mathrm{Ei} (- 2)]} \simeq 0.11
\end{eqnarray*}
a normalization factor and Ei the exponential integral function {\cite{AS65}}
.

The resulting size distribution function is peaked around $r_c \sim t^{1 / 2}$
and there are no domains with sizes larger than $2 r_c$ (Fig. \ref{fig:pdf}).

The physical meaning of (\ref{eq:separatrice}) can be understood by rewriting
the deterministic part of the equation of domain growth (\ref{eq:domain})
using $\rho$:
\begin{eqnarray}
  \frac{\gamma r_c^2}{\sigma}  \dot{\rho} & = & - \frac{\frac{\gamma r_0^2}{2
  \sigma t_0} \rho^2 - \rho + 1}{\rho}  \label{eq:dinamico}
\end{eqnarray}
The analysis of the fixed points of (\ref{eq:dinamico}) shows that when
condition (\ref{eq:separatrice}) is not satisfied, either the total domain
area grows to infinity, or shrinks to zero{\hspace{1pt}}{\footnote{See Refs.
{\cite{LP81,Bra94}} for the analogous discussion in the case of a locally
conserved field.}}. In both cases, the asymptotic condition (\ref{eq:kin2})
cannot be satisfied. Therefore, condition (\ref{eq:separatrice}) provides the
correct asymptotic distribution of domain sizes by selecting the separatrix
which divides those two extreme cases.

\begin{figure}[ht]
\centering
  \resizebox{7cm}{!}{\includegraphics{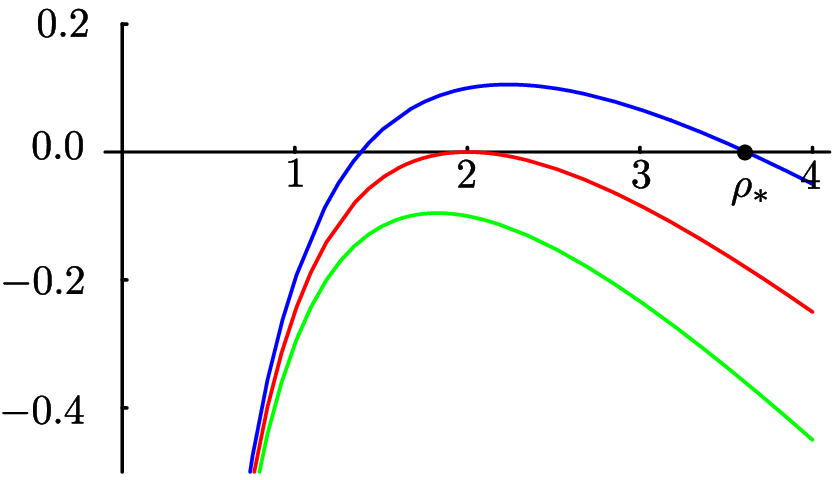}}
  \caption{Graph of the r.h.s. of (\ref{eq:dinamico}) for $\alpha \equiv
  \gamma r_0^2 / 2 \sigma t_0 = 0.2, 0.25, 0.3$. When $\alpha < 1 / 4$ Eq.
  (\ref{eq:dinamico}) has a fixed point $\rho = \rho_{\ast} > r_c$, which
  grows indefinitely with time, so that all domains grow and the total domain
  area grows to infinity. When $\alpha > 1 / 4$ all domains shrink to zero.
  The correct asymptotic behavior is found by selecting the separatrix between
  these two extreme cases.}
\end{figure}

\end{document}